\documentclass[]{spie}  

\usepackage{amsmath,amsfonts,amssymb}
\usepackage{graphicx}
\usepackage[colorlinks=true, allcolors=blue]{hyperref}
\usepackage{makecell}
\usepackage{multirow}

\title{Laying the Groundwork for the Development of the\\ Data Archive of the New Robotic Telescope}

\author[a]{Marco C. Lam}
\author[b]{Bovornpratch Vijarnwannaluk}
\author[b]{Pathompong Butpan}
\author[a]{Christopher M. Copperwheat}
\author[a]{Andrzej S. Piascik}
\author[b]{Utane Sawangwit}
\author[a]{Robert J. Smith}
\author[a]{Iain A. Steele}
\affil[a]{Astrophysics Research Institute, Liverpool John Moores University, IC2, Liverpool Science Park, 146 Brownlow Hill, Liverpool, UK L3 5RF}
\affil[b]{National Astronomical Research Institute of Thailand, 260 Moo 4, T. Donkaew, A. Maerim, Chiangmai, 50180 Thailand}

\authorinfo{Correspondence e-mail: c.y.lam@ljmu.ac.uk}

\begin{document} 
\maketitle

\begin{abstract}
The Liverpool Telescope has been in fully autonomous operation since 2004. The supporting data archive facility has largely been untouched. The data provision service has not been an issue although some modernisation of the system is desirable. This project is timely. Not only does it suit the upgrade of the current LT data archive, it is in line with the design phase of the New Robotic Telescope which will be online in the early-2020s; and with the development of a new data archive facility for a range of telescopes at the National Astronomical Research Institute of Thailand. The Newton Fund enabled us to collaborate in designing a new versatile generic system that serves all purposes. In the end, we conclude that a single system would not meet the needs of all parties and only adopt similar front-ends while the back-ends are bespoke to our respective systems and data-flows.
\end{abstract}


\section{INTRODUCTION}
\label{sec:intro}
A mutual interest of the National Astronomical Research Institute of Thailand~(NARIT) and the Liverpool Telescope~(LT)\cite{2004SPIE.5489..679S} group at the Astrophysics Research Institute at the Liverpool John Moores University is an upgrade to their data handling systems. At the time of writing, NARIT has not yet deployed a complete archive facility and LT is using 2000-era software which lacks flexibility. Looking to the future, we must also consider the needs of the joint operation (e.g. United Kingdom, Spain, Thailand and China) of the New Robotic Telescope~(NRT)\cite{2014SPIE.9145E..11C, 2016SPIE.9906E..3AC} which aims to be the World's largest and fastest response fully robotic telescope.

The archive project was funded by the UK Newton Fund delivered through the Science and Technology Facilities Council~(STFC) of the United Kingdom. The Newton Fund is a UK contribution to the official development assistance~(ODA) goals of the Organisation for Economic Co-operation and Development (OECD). It supports projects in specifically targeted geographic and subject areas with the aim to further sustainable development and improve welfare. Our project, {\sl Data-flow and Archiving for Robotic Operations Present and Future}, targets three of the 17 Sustainable Development Goals~(SDG) of the 2030 Agenda for Sustainable Development:

\begin{itemize}

\item{{\bf SDG\,4}: {\sl Ensure inclusive and quality education for all and promote lifelong learning.}}
\item[-]{NARIT has extensive schools and outreach activities to bring students into science and technology.}
\item[-]{Despite having a wide range of telescopes and data, NARIT need a data distribution system accessible by widely different users from primary school children to research physicists.}

\item{{\bf SDG\,8}: {\sl Promote inclusive and sustainable economic growth, employment and decent work for all.}}
\item[-]{Data processing, archiving and distribution facilitates industrial cooperation and enhances the impact of domestic innovation.}
\item[-]{Data fluency is critical to global knowledge based economy~(c.f.,\ agricultural or primary production economy).}
\item[-]{Astrophysics research is an attractive training ground for more widely applicable techniques and technologies.}

\item{{\bf SDG\,9}: {\sl Build resilient infrastructure, promote sustainable industrialization and foster innovation.}}
\item[-]{NARIT is an astronomical institution with a wide variety of research and engineering for both their own telescopes and external contracts.}
\item[-]{Skills needed for creation, operation and exploitation of a modern observatory are highly transferable to high-tech industries.}
\item[-]{NARIT has the telescopes, data and skills. The lack of mechanisms for data distribution are inhibiting Thailand's recognition in the international astrophysical community.}

\end{itemize}

\section{Database Management System}
\label{sec:dbms}
A database management system~(DBMS) is software to create and facilitate  access to databases. A DBMS provides a systematic way to insert, retrieve, update and delete data in a database. It serves as an interface between the database and the end user. It can provide concurrency, security, data integrity and uniform administration procedures. Most DBMSs can support a wide range of functions, for example, automated rollbacks, restarts, recovery, logging and auditing of activities. When data management is centralised, data can be protected and maintained more easily. Users from different locations can access a DBMS simultaneously in a controlled manner -- a DBMS can control the traffic and what the users can access.

\subsection{Choosing the right DBMS}
There are generally recognized to be four types\footnote{These are categorised by the nature of the data, not by the method of implementation, which include (not exclusively) flat file, column-oriented, document-oriented, object-oriented and key-value.} of DBMSs: Hierarchical, Network, Relational and Object-Oriented. They are each purposed for managing data with different characteristics. A key factor in the choice of DBMS is the nature of the data its structure.

\begin{itemize}

\item {\bf Hierarchical}
In a hierarchical DBMS, records contain information about the ``parent--child'' relationships like a tree structure. It is very fast and simple. Data follow a series of records, it collects all records together as a record type. These record types are equivalent to the tables in the relational model, with the individual records being the equivalent rows. To create links between these record types, the hierarchical model uses these type relationships. Hierarchical database can be accessed and updated rapidly because in this model structure is like as a tree and the relationships between records are defined in advance. However, this type of database structure is that each child in the tree may have only one parent, and relationships between children are not permitted. Adding a new field requires redefining the entire database. 

\item {\bf Relational}
Data in a relational database are stored in tables (i.e.\ relations) of rows\footnote{Also called {\it records} or {\it tuples}.} and columns\footnote{Also called {\it attributes}.}. Generally, each table represents one ``entity type'' -- the columns represent values attributed to that instance~(such as ``Fruit'' or ``Tree'') and the rows represent the instances of that type of entity~(such as ``Apple'' or ``Oak''). Each table has a key field that uniquely indicates each row, and these key fields can be used to connect one table to another.

\item{\bf Network}
Network databases are mainly used on large computers. When there are many connections between different types of data, network databases are considered more efficient. It also has a hierarchical structure, but a network database looks more like a cobweb of interconnected network of records -- each child can have more than one parent. Some data can naturally be modelled with multiple parents per child~(i.e.\ many-to-many relationships in data).

\item{\bf Object-Oriented}
Object-oriented databases are capable of handling any data type~(e.g.\ image, audio, video etc.), it is derived from the concept of object-oriented programming. They use small and reusable piece of software~(object) to store a piece of data and a method to interpret them. The benefits from using object-oriented databases are compelling. The ability to mix and match reusable objects provides incredible multimedia capability.
\end{itemize}

In observational astronomy, data are rarely hierarchical. This option can therefore be eliminated at the first instance. The fact that astronomical data are highly varied, for example, image, data-cube, radio data and tabulated data, implies that an object-oriented database may seem a good choice as it can handle them with ease. However previous experiments by the Sloan Digital Sky Survey (SDSS) with a then state-of-the-art object-oriented DBMS specifically written for astronomical use showed that the performance was not quite comparable to that of relational databases\cite{2004cs........3020T}. SDSS eventually decided to migrate their multi-terabyte object oriented system back to a relational DBMS\footnote{This was a huge volume at the time when the {\it largest commercialised} single hard disk was $\sim$\,$250$\,GB in 2003, compared with $\sim$\,$14$\,TB in 2018 at the time of writing}.

Both network and relational database management systems seem suitable for deploying an astronomical database. On one hand, most observational astronomical data are stored in FITS format which is standardised and static; on the other hand, they can vary in nature depending on the instrument and the science area. There is no one system that can efficiently manage a database in every way; in the commercial world many businesses rely on more than a single database for different tasks. Big data is the driving motivation to switch to using a network database, offering what traditional relational databases cannot do efficiently, for example, semantic query or handling linked data.

Although both systems show a lot of promising features, an audit of the real-world usage of the existing LT archive (see Section~\ref{sec:usage}) shows that searching through data using a network database would provide performance gain to only a small group of our users. In the following, only the most popular systems are investigated from the sustainability and maintenance point of view.

\subsection{Relational DBMS -- Structured Query Language~(SQL) database}
Relational databases almost exclusively use SQL as the query language for work and management. There are a number of different implementations of SQLs, but not all are compliant with the SQL Standards and many have incompatible language extensions. Hence most SQL query code is not portable among different databases implementations because they adopt slightly different syntaxes. Nevertheless, they are all very similar, most services improve the portability by limiting the use of functions to those that are mutual to all SQL variants. 

Most operations with SQL databases are ACID\footnote{{\bf A}tomicity, {\bf C}onsistency, {\bf I}solation and {\bf D}urability} compliant. For the extended features, one needs to check carefully which ones are and which are not. For example, despite PSQL supporting NoSQL functionality, operations with JSON tables remain ACID compliant; on the other hand, hash indexing lacked {\it durability} until v10.0. We provide a brief summary of MySQL and PSQL which were our selected candidate DBMSs:
\begin{enumerate}
\item {\bf MySQL} -- GPLv2\\
It is the most popular open source relational database, which also supports document store~(v5.7.12+) and key-value store~(v5.6+). The current LT archive is managed by MySQL. The main shortcoming is the query speed, particularly when searching data stored as spherical coordinates, such as equatorial Right Ascension and Declination. This can be overcome by adding indexes to the database, for example, B-Tree for ordered data and Dynamic Index Facility\cite{2010AdAst2010E..59N}~(DIF) for spatial data in a spherical polar coordinate system. DIF is currently an alpha development release.
\item {\bf PostgreSQL} -- PostgreSQL License\\
It is advertised as the relational database with high extensibility and standards compliance. It has similar bottleneck as MySQL, but the indexing technology on spatial query is substantially more mature. There are two available packages: Q3C and PgSphere which provide spatial indexing in a spherical polar coordinate system. Recent versions~(v9.2+) start to support document- and object-oriented data.
\end{enumerate}

\subsection*{Reasons to use a Relational Database}
\begin{itemize}
\item ACID compliance -- reduces anomalies and protects the integrity of your database by prescribing exactly how transactions interact with the database. Generally, network databases sacrifice ACID compliance for flexibility and processing speed.
\item Data are structured and unchanging -- if the data are consistent, there is no reason to sacrifice the high performance in relational search to support a few data types.
\end{itemize}

\subsection{Network DBMSs -- Not-only SQL~(NoSQL) database}
NoSQL databases are gaining popularity in recent years because of the rise of {\bf big data} and their scalability of data volume and flexibility with data type. NoSQL databases are often able to allow developers to execute queries without having to learn SQL or understand the architecture of their database system. Their performances outperform SQL databases in some areas significantly, for example, nearest neighbour search with a large data set. However, multi-rows/column operations are in general less efficient. The most popular open source NoSQL systems according to DB-engines.com are:
\begin{enumerate}
\item {\bf MongoDB} -- AGPL v3 License\\
It is a document-oriented database which is characterised by the schema-free organisation of data. Records do not need a uniform structure, columns can be arrays and records can have a nested structure. It supports ACID transactions by requiring all reads and writes are applied to the primary member by default. The overview on their webpage suggests to use PSQL for multi-row transactions, which is one of the most common types of transactions for an Astronomical database. Furthermore, it was most recently reported with 26,000 attacks in September 2017, adding to the 22,000 attacks in December 2016 due to the same security issue.
\item {\bf Apache Cassandra} -- Apache 2.0 License\\
Wide column stores database stores data in records with an ability to hold very large numbers of dynamic columns. Both the column names and the record keys are not fixed. A record can have any number of columns, so it can be viewed as a two-dimensional key-value store. It can achieve very high read/write performance and superior to many other NoSQL database in this sense. However, it lacks {\it consistency}, an update may affect one column while another affects the other, resulting in sets of values within the row that were never specified or intended.
\item {\bf Neo4j} -- GPL v3 License\\
Being a graph-oriented DBMS, it represents data in graph structures  as nodes and edges -- the relationship between nodes. They allow easy processing of data in that form, and simple calculation of specific properties of the graph, such as the distance between two nodes. It is reported to be an ACID-compliant transactional database. It is the most popular graph database according to DB-engines.com, but performance tests show that its performance is rather unremarkable for non-graph operations\footnote{https://www.arangodb.com/2015/06/performance-comparison-between-arangodb-mongodb-neo4j-and-orientdb/}.
\end{enumerate}

\subsection*{Reasons to use a Network Database}
\begin{itemize}
\item No limits on the types of data -- one can store any data together without having to define the ``types'' of data in advance. New types data can be added with ease.
\item Cloud-based storage -- an potential cost-saving solution, but it requires data to be easily spread across multiple data centres.
\item Rapid development -- modifying a relational database can lead to a lot of down time. In contrast, processing network database data requires little overhead.
\end{itemize}

\section{Search Engine}

Search engines are databases optimised toward searching and performing analytics. They possess the ability to perform more complex queries than most DBMSs, such as fuzzy full-text search, spherical coordinate search, advanced aggregations, and relevancy weighting. These queries require additional indexing and they usually exist as third-party plug-ins.

In a relational DBMS, data are stored in a table-like structure. This structure and ACID transactions gives relational DBMS its famed reliability. Search engines are designed to cope with very large amounts of data and hence have to be scalable. Therefore, search engines generally follow a NoSQL like structure. For example, Elasticsearch (ES) keeps records in a file-like format which consists of key-value pairs. The terms ``file-like'' or ``document oriented'' stem from the conceptual manner of indexing and handling input documents as individual entities. It should not be taken to imply the data are stored as their original physical files.  The index exists as a proprietary binary search engine file, just as it would in a relational database.

\verb+JOIN+ is a common database operation that is not available in search engines. It is common to de-normalize the data into a single record before indexing into a search engine, so a single record will hold all information related to it. This is different from how a relational DBMS denormalises the data -- data are broken down and stored in multiple tables. For example, an observatory database may contain tables of users, observations, filters, instruments and exposures connected by cross-reference keys that allow those tables to be {\tt JOIN}ed. This is done to prevent storage and maintenance of large quantities of identically duplicated data. A search engine does not suffer from the same duplication overhead problems as a relational database so may simply hold records of {\sl `observations'} which include all associated exposures, details of the instruments and filters, and details of the observer who ordered it. Hence, \verb+JOIN+ would not be required since all data exist in the record itself.

It is worth noting that the distinction between search engines and databases is becoming increasingly blurred. Many relational DBMS and NoSQL databases have developed many of the same capabilities as search engines. PSQL and MySQL have already implemented full-text search capabilities, result caching, and geolocation data-types. Therefore, DBMS such as PSQL can be used in place of search engines by creating a database containing the searchable fields. The choice of either an off-the-shelf search engine or a relational DBMS is mostly a matter of preference and ease of implementation with the existing/available infrastructures.

\section{Usage Audit and User Requirement Survey}
\label{sec:usage}

In order to understand what search fields a new data archive should provide, we conducted a usage audit with the existing LT data archive and a user requirement survey with the scientists at NARIT.

\subsection{LT Data Archive Usage Audit}
In May 2017 we started saving all archive search logs rather than just pruning out what we thought would be useful. From that date we can therefore look back and derive statistics on all searches performed. 

Up to April 2018, the complete search logs span over 355 days, with 6,593 searches, averaging to 19.68 searches per day. These have been filtered down to only those queries that look like someone genuinely trying to extract something from the archive. I.e., they exclude queries run as part of automated internal maintenance and operational tasks. 

The statistics exclude National School Observatory~(NSO). The NSO is an important and substantial contributor to archive usage; the archive currently contains over 54,000 schools' observations and query traffic exceeds 600 per day. However these queries use a different interface and are all performed using a priori known search keys, so contain no useful information about how the data are being use.

The statistics also do not include a daily search and report that is automatically generated for the Gaia Ground Based Optical Tracking~(Gaia-GBOT) project since like the NSO data, that is a preset query, always run the same way. Notwithstanding discussion below, we consider the ability to automate regular queries like this to be an important facility for any modern observatory archive. 

A detailed breakdown of what query options are actually being used is given in Table~\ref{tab:audit}. The following are some highlights and comments:
\begin{itemize}
\item About half of all queries were constrained on equatorial coordinates.  Galactic coordinates are offered but were virtually unused (0.05\,\%).
\item Of order one third specify the desired target by name rather than coordinates. 
\item About half of all queries name a specific instrument.
\item 561 out of 6,593 (8.5\,\%) are instances of repeated searches suggesting a user might have a canned query they submit many times. Most users appear to be using the archive search tools only occasionally and interactively though this may just reflect the fact that the current interface is not simple to automate. 
\item The most commonly repeated query is 71 instances for a particular (\verb+RA+, \verb+Dec+) position. The second most common is 42 searches on a particular NGC target by name, followed by 33 instances of a particular supernova by name and then 25 instances of search of a specific \verb+username+.
\item A fairly small 7.3\,\% of searches made use of the output result formatting options.
\item Many searchable parameters which would appear observationally important (seeing, airmass, binning, spectral resolution) are not being widely used. This appears to mean that users are requesting everything available on a particular source and applying second tier selection functions for themselves. What is not clear is whether or not they are only doing that due to inflexibility of the current interface.
\item 2.2\,\% of searches asked for the raw data as well as reduced showing most users are very happy to rely on the automated pipeline data reductions.
\item Over 30\,\% of searches requested only data that were outside their proprietary period. Though we cannot identify the users performing these queries, it seems likely this is data mining by people other than the original principle investigator.
\end{itemize}

\begin{table}[h!]
\caption{The top search criteria in each of the three categories: Observational, Instrumental and Scheduling. (a) There were 2,125 searches for 655 unique object names. The top 20 unique names account for 27\,\% of name searches: 14\% are NGC objects, 22\,\% are Messier objects and 1\% are ``Moon". (b) By instrument name: 47\,\% of queries specified an instrument and of those 65\,\% IO:O, 13\,\% Ringo, 8\,\% RATCam, 6\,\% SPRAT, 4\,\% IO:I, 4\,\% FRODOSpec, 4\,\% RISE, 1\% LOTUS and $<$1\% Meaburn, SupIRCam, Thor. The numbers do not add up to 100 because a single query may ask for several instruments. (c) Of which, 60\,\% Science, 32\,\% Standard, 6\,\% Arc, 2\,\% Lamp-Flat.
}
\label{tab:audit}
\centering
\hfill \\
\begin{tabular}{|c|c||c|c||c|c|}
\hline
{\bf Observational} & {\bf \%} & {\bf Instrumental} & {\bf \%} & {\bf Scheduling} & {\bf \%} \\ \hline
Equatorial (R.A., Dec.) & 52.43 & Instrument$^{b}$    & 47.13 & Non-proprietary only       & 31.35 \\ \hline
Object Name$^{a}$       & 32.12 & Filter              & 27.71 & Proposal ID                & 10.10 \\ \hline
Calendar Date           & 10.77 & Integration Time    &  4.16 & User ID                    &  6.14 \\ \hline
Seeing                  &  2.44 & Instrument Mode     &  1.40 & Observation Type$^{c}$     &  6.10 \\ \hline
Airmass                 &  0.20 & Binning             &  0.71 & Data Reduction Level
&  4.29 \\ \hline
MJD                     &  0.12 & Spectral Range      &  0.42 & Filename                   &  2.84 \\ \hline
Galactic (l, b)         &  0.05 & Spectral Dispersion &  0.09 & TAG ID                     &  2.14 \\ \hline
 ---                    &  ---  & Spectral Resolution &  0.00 & Observation ID             &  0.68 \\ \hline
 ---                    &  ---  & ---                 &  ---  & Group ID                   &  0.49 \\ \hline
\end{tabular}
\end{table}

\subsection{NARIT User Requirement Survey}

A data archive is an important infrastructure that NARIT currently lacks. Researchers use either portable hard disk or Secure Copy Protocol~(\verb+scp+) to transfer data from the file repository to their own computers. This method is sufficient in most cases given that the observers are usually the only users of the data. However, it is imperative to develop an archive system for search and retrieval of data. We started with a basic questionnaire to understand the query behaviours of our users, who also routinely access other data archives as part of their research work. 

The questionnaire is divided into three sections, user information, science cases, and data archive experience. In the last section, respondents were asked about their previous experience with other data archives. Some questions are concerned about the format of the data files requested, other questions are concerned about additional features that would be useful to them. A few example requests include: automatic data retrieval for users asking for their own data and, observation statistics and related time series data associated to a particular observation. The most important questions in designing a user interface is how users search for data. To find out, respondents were presented with a list of search fields which are commonly given in archives and were asked to identify which fields they normally use. Among the 7 respondents, 6 search by the target name, 5 by the calendar date, instrument, coordinates, and 4 by the filters and the name of the telescope. When the researchers are asked to identify their most preferred search field, i.e.,\ if there is only one search field available with the database, which one would they choose: 43\% prefer using the object name, 29\% would search by the coordinates, 14\% by the name of the observer and 14\% by the calendar date. 

It is difficult to draw concrete statistical usage trends due to the small number of respondents. However, the questionnaire results follow a similar trend in the LT archive audit which shows that a majority of users search with object coordinates and object names. Given the two independent surveys coming up with similar results, we decided these commonly queried keys should exist as the baseline-requirements of the front-end interface in the future data archive.

\section{Feasibility Test and Search Optimisation}

From the analysis of the usage audit and user survey, it is obvious that the majority of the queries would require partial or full text search~(P/FTS) on one or more fields, and by the spatial location. On the other hand, numeric fields are not commonly used as search criteria.

\subsection{Feasibility Test}
Eight use cases were tested with PostgreSQL~(PSQL) and Elasticsearch~(ES) at the beginning of our prototyping exercise~(see Table~\ref{tab:querytime} and Section~\ref{sec:prototype}). Both PSQL and ES were individually optimised by indexing the fields with the most appropriate types of index~(see below).

\begin{table}[h!]
\caption{Median query time of 100 repetitions for the eight use cases compared between PSQL and ES, sorted by the query time in PSQL. The time is reported in milliseconds.}
\centering
\hfill \\
\begin{tabular}{|c|c|c|}
\hline
\multirow{2}{*}{\bf Use case} & \multicolumn{2}{c|}{\bf Execution Time } \\ \cline{2-3}
              &   {\bf PSQL} & {\bf ES} \\ \hline 
PTS on \verb+objectname+ with \verb+m42+ & $<$1 & 3 \\
2' radius cone search at (00:00:00, +10:00:00) & $<$1 & 30 \\
10' radius cone search at (00:00:00, +89:59:00) & $<$1 & 3 \\
20' radius cone search at (12:00:00, +50:00:00) & $<$1 & 5 \\
PTS on \verb+objectname+, \verb+userid+, \verb+obsid+, \verb+groupid+, \verb+propid+ and \verb+tagid+ with \verb+1058+ & 97 & 187 \\
Total exposure time for each of the 7 instruments in 2015 & 3816 & 110 \\
Calculate the mean and RMS seeing in 2014 & 4660 & 72 \\ \hline
\end{tabular}
\label{tab:querytime}
\end{table}

\subsection{Database Index}
A database index is a data structure that improves the search speed at the cost of additional write time and storage space. Indexes can be created on one or more fields of a database to locate data without having to go through every item in the database. Different types of indexes usually cannot be combined, a second index can only be applied, if possible, to the subset of the database filtered by the first index. With PSQL, the extension \verb+btree_gin+ can be enabled to allow the joining of B-Tree and GIN index. The following describes only the indexes we have used in the prototype, and we have turned on the \verb+btree_gin+.

\subsection*{B-Tree Index}
A self-balanced tree (B-tree) is a generalised binary search tree where a parent node can have more than two child branches (whereas in a binary tree, each node has exactly two child branches). This can provide a high performance in reading and writing large blocks of sorted data. Because of the nested search nature, it has a logarithmic time complexity. We have applied this index to the seeing, airmass and exposure time. 

\subsection*{Generalised Inverted Index~(GIN)}
It is designed for handling cases where the items to be indexed are composite values, and the queries search for values that appear within the composite items. In the case of using documents as an item, the queries would be to search for all documents containing specific words. Despite it being most commonly used for matching words or phrases in documents, it can be useful to any dataset that exhibits one-to-many mapping. By looking up the key in the GIN index, the system can immediately locate the requested data. In PSQL, GIN is designed for text search. In order to apply it to other data types~(e.g.,\ int4, float8, date, varchar, text etc.) the \verb+btree_gin+ extension has to be installed. In order to perform text search efficiently, a column of text search vector has to be added to the table~(see below).

\subsection*{Spherical Spatial Index}
When data carry 2-dimensional spatial information a nearest neighbour search will have quadratic time complexity. Conventional index types cannot efficiently handle spatial queries such as finding distances between all points, or selecting all points within a spatial area of interest. The native PSQL spatial index implements the R-tree, which is a tree of minimum-bounding-rectangle. It is extremely efficient for the said purpose, Google Maps uses a variant of such an implementation, although the precise details are a commercial secret. This index type, however, cannot be directly applied to a spherical polar system (i.e., the celestial sphere) because it cannot detect the periodic boundary and it completely fails near the poles. Extension packages have to be installed in order to work in a spherical polar system

\begin{itemize}
\item pgSphere~\cite{2004ASPC..314..225C}
\item[] This package uses GiST~(Generalized Search Tree) to implement R-Tree on spherical objects. The search algorithm divides the sample on a 2D plane into component minimum bounding rectangles~(MBRs) hierarchically so that when searching for an object, one only needs to go through a subset of the MBRs hence providing significant performance gain in nearest neighbour or other distance related searches.
\item Q3C~\cite{2006ASPC..351..735K}
\item[] Q3C is short for Quad-Tree-Cube. It is a pixelisation scheme that projects the 6 faces of an inscribed cube onto a spherical surface. Quad-trees divides each node of a tree data structure into exactly four children hierarchically. They were independently constructed on each face of the cube to pixelise the surface into equal areas. However, the radial projection of the cube surface onto a spherical surface does not preserve the equal-area property.
\item Geohash

A geohash is an alphanumeric string which represents geolocation coordinates. The system works by subdividing Earth coordinates into rectangular cells which are represented by each char in the string. The same process is repeated for each cell to increase precision and a new char is added to the end of the string. In other words, the longer the string, the more precise the coordinate will be. For two strings, the longer the sub-strings that share the same prefix, the closer they are together. Geohash does suffer from some edge cases but workarounds have been created and implemented by most services.

\end{itemize}

\subsection*{Full/Partial Text Search}
\label{sec:fts}
In PSQL, text search vector can be created on a field to speed up F/PTS. It is a sorted list of distinct lexemes, which are words that have been normalised to merge different variants of the same word. This introduces extra overhead to generate the \verb+tsvector+, but since this can work on top of the GIN index and be combined with most other indexes, it can provide significant performance gain.

In Elasticsearch (ES), FTS relies on searching an inverted index. The ingested text is broken down and normalised into terms that are kept in the inverted index. Searches scan through the index for matching terms and those documents which contain most matches are returned as results with the highest relevancy. ES possesses many full-text query methods available out-of-the-box, such as normal match queries, exact-match queries (term queries), and phrase queries. Fuzzy matching and wild card queries can also be used.

\subsection{Database Partition/Shard}
We have not introduced any database partitioning to our PSQL test because it only provides performance gain when a specific column is heavily used. However, it is worth mentioning that when performing light curve analysis of SkyCamT\cite{2014SPIE.9152E..2LB} data from the LT\footnote{http://telescope.livjm.ac.uk/TelInst/Inst/SkyCam/}, the partitioning on the Object ID in the measurement table provided an increase in performance by about one to two orders of magnitude depending on the number of epochs available with the objects.

On the other hand, ES was designed to be a distributed system, it can be distributed over multiple servers which hold different sections of an ES index. This will prove useful when the ES index becomes too large for a single node to hold or too large to efficiently search. In such a cluster environment, ES distributes different sections (shards) of the index to each node and creates replicas of each shard as a redundancy in case any node fails to start up. This whole process is carried out by ES and users are usually unaware of it. The user may specify the number of shards when creating the index. However, there is no official recommendation on how to optimize the number of shards for an index. Hence, it is left to the engineers to work out the optimal number of shards. As for the number of replicas, the recommendation is to always have two replicas per shard.

Sharding an ES index allows queries to be performed in parallel within each ES node. It also hardens the index against failure with replicas for each shard. Replicas are distributed to always be on a different node from their shard counterpart. If a node fails to start up, ES automatically promotes the replicas to shards. In the case of indexing, ES only allows indexing to be done by a single ``primary shard'' in order not to cause writing conflicts in the index. In the case that the primary shard is down, ES automatically promotes another shard to serve as a primary shard.

\section{Prototype}
\label{sec:prototype}
Two prototyping projects were carried out simultaneously over a 10-day period, one with PSQL and the other one with ES. PSQL is a closer fit to the needs of LT operations where file formats are standardised and it only needs to serve data for a single optical observing facility. ES is designed to cope with a wider range of source facilities, which at NARIT range from optical through infrared to radio data, all with different metadata information.

For testing, we cloned the complete FITS header catalogue from the LT archive~(15 years of data) as the test dataset. Through which the ES were tested over a large established dataset to understand its capabilities, especially ones that are well understood with a relational database. Independently, PSQL was indexed with a mix of B-Tree, GIN and pgSphere to test the potential performance gain that could bring to the query service. These were combined with a front-end technology demonstrator developed mostly, but not exclusively, in JavaScript, AngularJS, jQuery and D3.js\cite{2011-d3}, where query forms were created for each of the prototypes. Operation logs and science data were combined and ingested with Logstash and displayed with Kibana. Table~\ref{tab:tech} shows the software we have adopted in the current archive in operation and the stacks of the prototypes.

\begin{table}[h!]
\centering
\caption{Summary of the technologies (to be) employed by the current and the proposed data archives.}
\label{tab:tech}
\hfill \\
\begin{tabular}{|l|c|c|c|}
\hline
{\bf Technology}   & {\bf \makecell{Current\\ LT Archive}} & {\bf \makecell{Proposed\\ New LT Archive}}      & {\bf \makecell{Proposed\\ NARIT Archive}} \\ \hline
Markup Language    & HTML4                    & HTML5                              & HTML5 \\ \hline
Style Sheet        & CSS2                     & CSS3                               & CSS3 \\ \hline
Dynamic Display    & cgi-bin                  & \makecell{AngularJS\\ jQuery}      & \makecell{AngularJS\\ jQuery} \\ \hline
Web Server         & Apache HTTP Server       & \makecell{Node.js\\ Express.js}    & \makecell{Node.js\\ Express.js} \\ \hline
Web-log Formatter  & N/A. Flat ASCII files    & T.B.D.               & \makecell{morgan-json\\ Logstash} \\ \hline
Command-line Maintenance & \makecell{sh\\ PHP\\ DBMS native tools} & \makecell{sh\\ PHP\\ DBMS native tools} & N/A \\ \hline
 Web-based dashboard  & Some minimal PHP      & JavaScript         & Kibana \\ \hline
Backend Language   & C       & JavaScript                         & JavaScript \\ \hline
Package Management & ---                      & npm                                & npm \\ \hline
Database           & MySQL                    & PostgreSQL                         & Elasticsearch \\ \hline
Index              & B-Tree                   & \makecell{B-Tree\\ GIN\\ pgSphere} & \makecell{Geohash\\ Inverted-index} \\ \hline
VO-compatible      & Yes (Not maintained)     & Yes & Planned \\ \hline
\end{tabular}
\end{table}

\section{Conclusion}

In our exploratory work, we have tested two new stacks with PSQL and ES. Either satisfies most use cases in searching and monitoring data. In any system, even in ES which can accept irregular data formats and headers, the completeness of metadata is necessary for the ingested data to be searched. The most essential are coordinates and unique IDs. This requires additional data processing which is the cost coming with the flexibility of ES. The choice of the data distribution service should focus on the features one is providing and the extent of work required to maintain the system.

In the future, it is worth understanding the feasibility in using cloud storage for facilities of different sizes. The data stored must be General Data Protection Regulation~(GDPR)\footnote{https://www.eugdpr.org/} compliant and the interface should ideally be IVOA\footnote{http://www.ivoa.net/} compliant by design. 

\acknowledgments 
This project was made possible by the financial support of the Newton Fund, delivered through the UK Science and Technology Facilities Council.

The Liverpool Telescope is operated on the island of La Palma by Liverpool John Moores University in the Spanish Observatorio del Roque de los Muchachos of the Instituto de Astrofisica de Canarias with financial support from the UK Science and Technology Facilities Council.

National Astronomical Research Institute of Thailand~(Public Organisation) operates at the city of Chiang Mai, Thailand. Under the financial support of the Ministry of Science and Technology of Thailand.

\bibliographystyle{spiebib} 
\bibliography{lt-data-archive} 

\end{document}